\documentclass[prc,twocolumn,twoside,showpacs]{revtex4}

\usepackage{hyperref}
\usepackage{amsmath}
\usepackage{graphicx}

\begin{document}

\title{Microscopic Calculation of Fusion Cross-Sections}

\author{Thomas Neff}
\email[Electronic address: ]{t.neff@gsi.de}
\author{Hans Feldmeier}
\author{Karlheinz Langanke}
\affiliation{Gesellschaft f{\"u}r Schwerionenforschung mbH,
  Planckstra{\ss}e 1, 64291 Darmstadt, Germany}

\date{\today}

\newcommand{\bra}[1]{\big< {#1} \big| }
\newcommand{\ket}[1]{\big| {#1} \big> }
\newcommand{\braket}[2]{\big< {#1} \big| {#2} \big> }
\newcommand{\matrixe}[3]{\big< {#1}\, \big| {#2} \big| {#3} \big> }
\newcommand{\fm}{\ensuremath{\text{fm}}}
\newcommand{\MeV}{\ensuremath{\text{MeV}}}
\newcommand{\isotope}[2]{\ensuremath{{}^{#1}\mathrm{#2}}}

\begin{abstract}
  A microscopic calculation of cross sections for fusion of oxygen
  isotopes \isotope{16}{O}, \isotope{22}{O} and \isotope{24}{O} is
  presented. Fermionic Molecular Dynamics wave functions are used to
  describe the fully antisymmetrized and angular momentum projected
  nucleus-nucleus system. The same effective nucleon-nucleon
  interaction is used to determine the ground state properties of the
  nuclei as well as the nucleus-nucleus interaction. From the
  microscopic many-body wave function the corresponding wave function
  for the relative motion of two point-like nuclei is derived by a
  method proposed by Friedrich which leads to a local effective
  nucleus-nucleus potential. Finally the Schr{\"o}dinger equation with
  incoming wave boundary conditions is solved to obtain the
  penetration factors for the different partial waves. With these the
  S-factor for the fusion process is calculated. A good agreement with
  experimental data is obtained for the
  \isotope{16}{O}-\isotope{16}{O} cross section. Much enhanced cross
  sections are found for the neutron-rich oxygen isotopes.
\end{abstract}


\maketitle

\section{\label{sec:introduction} Introduction}

Heavy-ion fusion reactions play important roles in various
astrophysical objects. For example, reactions induced by the fusion of
two \isotope{16}{O} nuclides are the main energy source of hydrostatic
oxygen burning \cite{rolfsrodney,woosley02}.  On the other hand pycnonuclear
fusion reactions of neutron-rich nuclides are expected to occur in the
crust of accreting neutron stars in binary systems, when the ashes of
x-ray bursts are pressed to high densities and a series of electron
captures has transformed the matter neutron-rich \cite{haensel90,haensel03}.
Electron captures and pycnonuclear fusion generate energies which can
affect the dynamics and evolution of the neutron star crust
\cite{bildstein98}.

To determine the relevant fusion cross sections is quite challenging.
While for the \isotope{16}{O}+\isotope{16}{O} reaction data exist at
energies below the Coulomb barrier \cite{spinka72,hulke80,wu84,thomas86}, it
is yet impossible to determine the cross sections at the most
effective energies of hydrostatic oxygen burning ($E \approx 2$ MeV).
Hence it has been the standard procedure to extrapolate the measured
cross sections down into the energy region of astrophysical interest
usually based on phenomenological barrier penetration models
\cite{caughlan88}. Furthermore, a calculation of the subbarrier
\isotope{16}{O}+\isotope{16}{O} fusion cross sections performed on the
basis of the Hartree-Fock model agrees with the data quite well
\cite{reinhard84}. The situation is significantly more uncertain for the
determination of the fusion cross sections for the neutron-rich
nuclei. Here no data exist and the cross sections have to be derived
solely from theoretical models (e.g. \cite{gasques04}).  Furthermore
neutron-rich nuclei develop neutron densities with long tails which
lead to strongly enhanced fusion probabilities, but are very difficult
to systematically describe by parametrized potentials \cite{gasques04}.

Obviously it is desirable to base the required extrapolations on
theoretical models which are as reliably as possible; i.e.  which are
accounting for the relevant degrees of freedom of the reaction. In
recent years powerful many-body approaches have been developed
(Green's Function Monte Carlo \cite{wiringa00}, No-Core Shell Model
\cite{navratil98}, Fermionic Molecular Dynamics (FMD)
\cite{fmd00,fmd04,hoyle07}) which allow now a quantitatively correct
description of light nuclei based on the solution of the many-body
problem for realistic nucleon-nucleon interactions. The next step
should be to take advantage of these advances in nuclear structure
physics for a more realistic and reliable description of low-energy
reaction problems.  First attempts have been presented in
\cite{arriaga91,navratil06} where microscopic structure information
derived from the GFMC or no-core shell model has been incorporated
into potential model descriptions of low-energy d+d and p+$^7$Be
fusion. Here we will present a different approach, based on the FMD
model, which allows for a consistent description of bound, resonant
and scattering states from the same many-body Hamiltonian derived from
a realistic nucleon-nucleon interaction. In particular, our model
describes consistently the internal structure and properties of the
asymptotic reaction partners and the relevant degrees of freedom of
the collision process.  This is particularly important for
astrophysically relevant reactions including nuclei far from
stability, as their description is usually not constrained by data
yet, but it is expected to depend sensitively on nuclear structure
subtelities like the formation of neutron halos or skins. While the
FMD approach to nuclear reactions is quite general and can be applied
to scattering, capture and transfer reactions, we will here present a
first application which deals with subbarrier fusion of oxygen
isotopes.

We introduce the FMD many-body model in Sec.~\ref{sec:fmducom}. The
calculation of the microscopic nucleus-nucleus potential and the
transformation into an effective potential for point-like nuclei is
performed in the language of the well known Generator Coordinate
Method (GCM) and Resonating Group Method (RGM) and is presented in
Sec.~\ref{sec:collective} and \ref{sec:potentials}. In
Sec.~\ref{sec:fusion} the fusion cross sections are calculated by
solving the Schr{\"o}dinger equation using incoming wave boundary
conditions. In Sec.~\ref{sec:adiabatic} we discuss adiabatic effects
before summarizing in Sec.~\ref{sec:summary}.

\section{\label{sec:fmducom} Many-body model}

In the Fermionic Molecular Dynamics \cite{fmd00,fmd04} model the
many-body states are described by Slater determinants 
\begin{equation}
  \label{eq:fmdslaterdet}
  \ket{\psi} = \mathcal{A} \biggl\{ \ket{q_1} \otimes \ldots \otimes
    \ket{q_A} \biggr\}
\end{equation}
using Gaussian wave packets for the single-particle states
\begin{equation}
  \label{eq:spstate}
  \braket{\vec{x}}{q}= \exp \biggl\{ -
  \frac{(\vec{x} -\vec{b})^2}{2a} \biggr\} \otimes
  \ket{\chi^\uparrow, \chi^\downarrow} \otimes \ket{\xi} \: .
\end{equation}
The complex parameter $\vec{b}$ encodes the mean position and the mean
momentum of the wave-packet. In this paper the width parameter $a$ is
chosen to be identical for all single-particle states. The spins can
orient freely, the isospin is fixed to describe either a proton or a
neutron. The ground state wave function is obtained by minimizing the
intrinsic energy with respect to the single-particle parameters $a$,
$\vec{b}_i$ and $\chi_{i}$. For \isotope{16}{O}, \isotope{22}{O} and
\isotope{24}{O} we obtain configurations that are identical to
closed-shell harmonic oscillator configurations with a fully occupied
$0d_{5/2}$ shell in \isotope{22}{O} and a fully occupied $1s_{1/2}$ shell 
in \isotope{24}{O}.

We use a nucleon-nucleon interaction that is derived from the
Argonne~V18 interaction by explicitly introducing the short-range
central and tensor correlations into the many-body system by means of
the Unitary Correlation Operator method \cite{ucom03,ucom05} in a
two-body approximation. The correlated interaction has been used
successfully in \emph{ab initio} calculations using many-body
perturbation theory and the no-core shell model \cite{ucom06}. As the
simple FMD wave functions are not able to describe the medium- to
long-range tensor correlations we use a phenomenological modification
of the two-body interaction as introduced in \cite{fmd04}. Here an
additional momentum-dependent two-body part is used and fixed to
reproduce the binding energies and radii of \isotope{4}{He},
\isotope{16}{O} and \isotope{40}{Ca}. An additional spin-orbit term is
tuned to the binding-energies of \isotope{24}{O}, \isotope{34}{Si} and
\isotope{48}{Ca}.

\begin{table}
  \begin{tabular}{lrrr}
    & \isotope{16}{O} & \isotope{22}{O} & \isotope{24}{O} \\
    \hline\hline
    $E_b$ / MeV (Experiment)          & -127.62 & -162.03 & -168.48 \\
    $E_b$ / MeV (FMD)                 & -120.07 & -153.88 & -158.75 \\
    $r_\text{matter}$ / fm (Experiment) & 2.54(2) & 2.88(6) & 3.19(13) \\
    $r_\text{matter}$ / fm (FMD)        & 2.50 & 2.80 & 2.95 \\
  \end{tabular}
  \caption{Ground state energies and matter radii calculated
    with the FMD wave functions. 
    The experimental matter radii are taken from \cite{ozawa01}.} 
  \label{tab:groundstates}
\end{table}

\begin{figure}
  \includegraphics[width=\columnwidth]{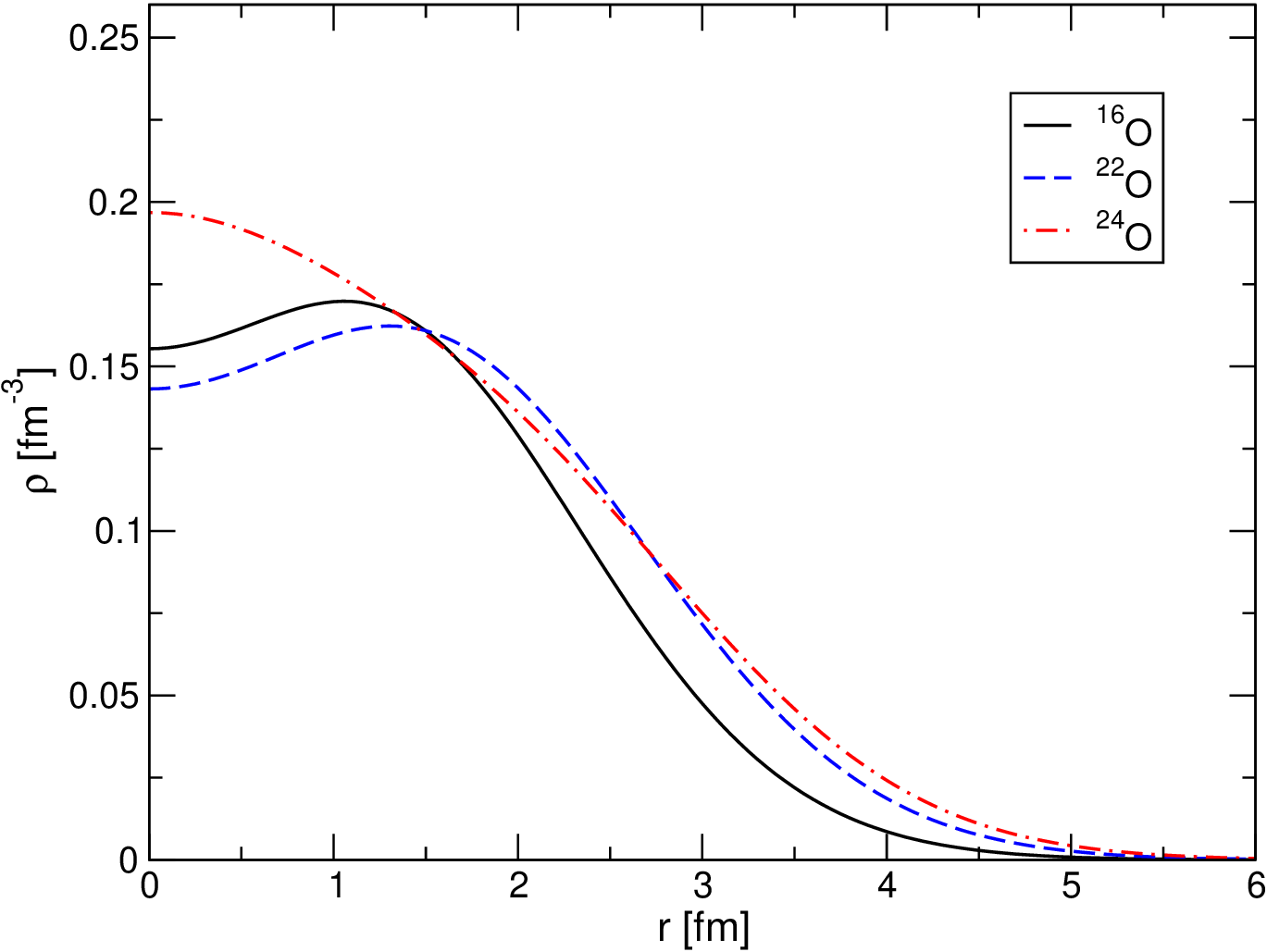}
  \caption{Matter densities for \isotope{16}{O}, \isotope{22}{O}
  and \isotope{24}{O}. Finite size of protons and neutrons is taken
  into account.}
  \label{fig:densities}
\end{figure}

\begin{figure}
  \includegraphics[width=\columnwidth]{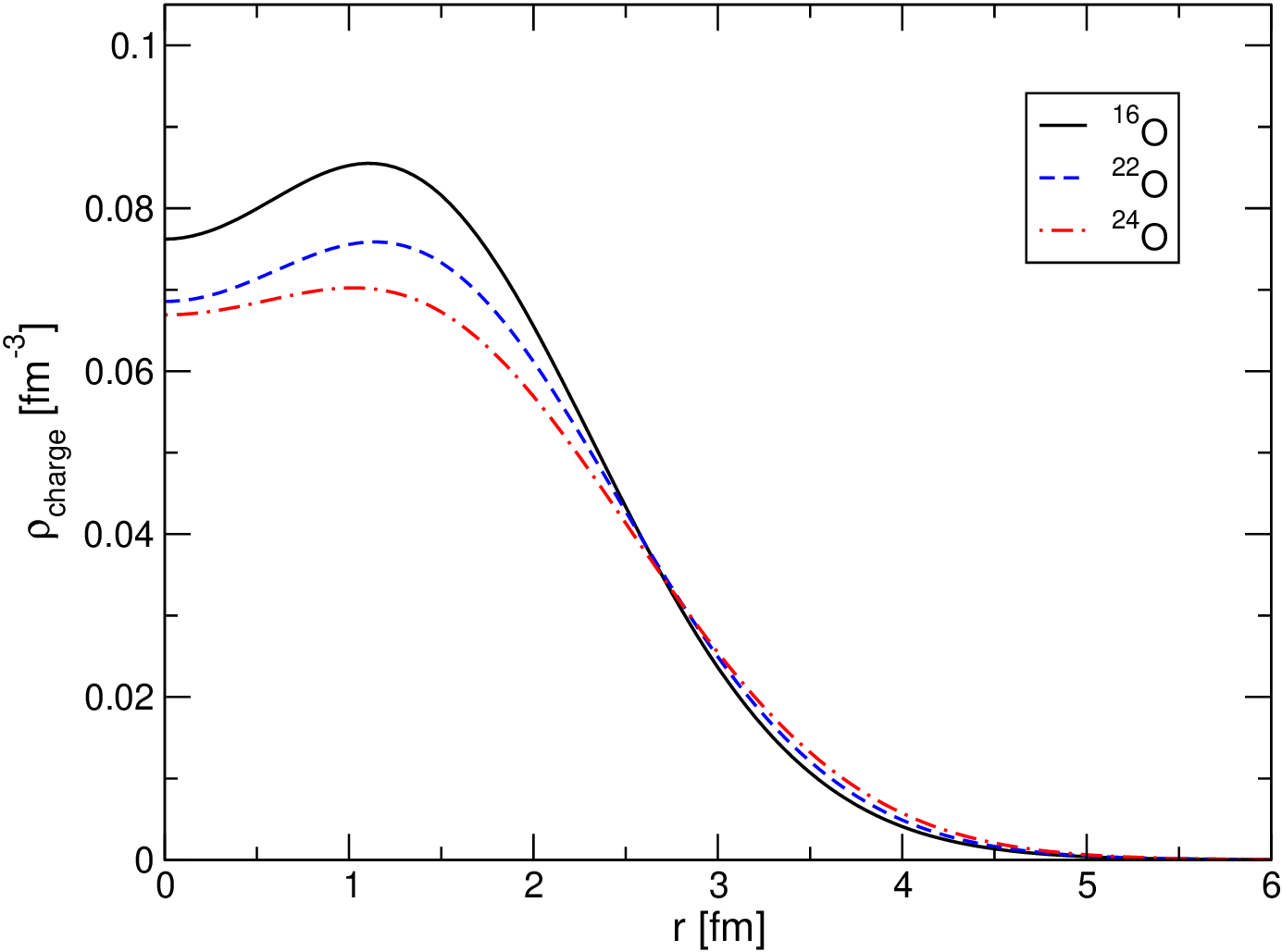}
  \caption{Charge densities for \isotope{16}{O}, \isotope{22}{O}
  and \isotope{24}{O}. Finite size of protons and neutrons is taken
  into account.}
  \label{fig:chargedensities}
\end{figure}

The calculated binding energies and matter radii are shown in
Tab.~\ref{tab:groundstates}. The ground state densities are shown in
Figs.~\ref{fig:densities}, \ref{fig:chargedensities}.  An improved
description of the ground states is possible in a variation after
projection approach. But as the intrinsic states would then no longer
be angular momentum eigenstates, the calculation of the
nucleus-nucleus potential would no longer be feasible. As the density
distribution remains almost unchanged we expect no significant effect
on the nucleus-nucleus potentials and therefore on the fusion cross
sections.


\section{\label{sec:collective} Wave functions for relative motion}

\subsection{GCM wave functions}

\begin{figure}
  \includegraphics[width=\columnwidth]{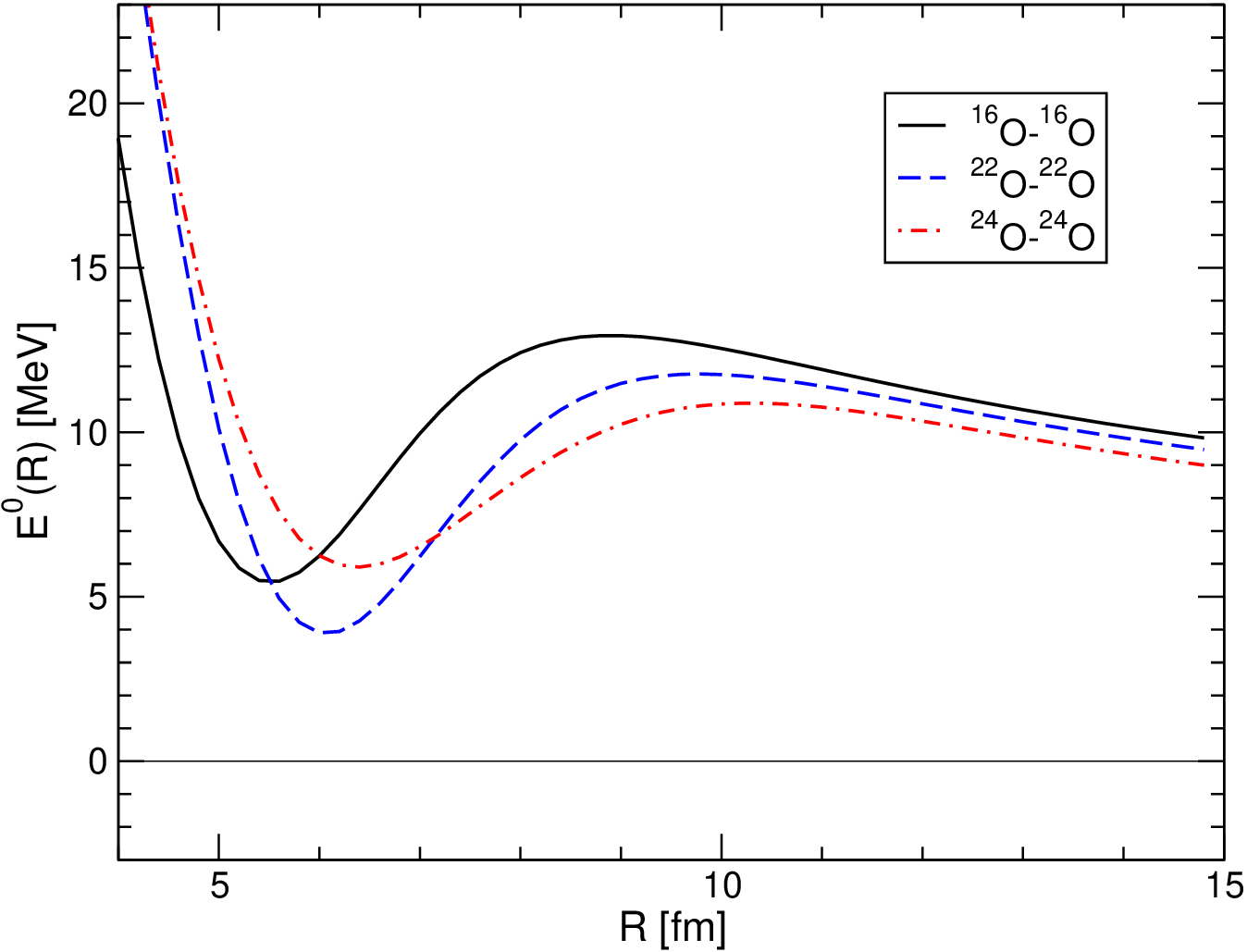}
  \caption{GCM energy surfaces for $L=0$. The energy surfaces are
    different even asymptotically because of different localization
    energies in the relative motion.}
  \label{fig:gcmsurfaces}
\end{figure}

To calculate the microscopic nucleus-nucleus energy surfaces we make
use of the fact that the FMD wave functions can be easily moved and
rotated. We construct Generator Coordinate Method (GCM) wave functions
using the normalized FMD ground state wave functions $\ket{\psi}$ and
the distance $\vec{R} = R \vec{e}_z$ between the origin of the slater
determinants as the generator coordinate. $R$ is assuming values from
$0.5\:\fm$ up to $15\:\fm$ in steps of $0.5\:\fm$
\begin{equation}
  \label{eq:gcmbasisstate}
  \ket{\Psi(\vec{R})} = \mathcal{A} \left\{ \ket{\psi(-\vec{R}/2)}
    \otimes \ket{\psi(+\vec{R}/2)} \right \} \: .
\end{equation}
We have to project these states on angular momentum with the
projection operator \cite{ringschuck}
\begin{equation}
  P^J_{MK} = \frac{2J+1}{8\pi^2} \int d\Omega \:
  D^{J^\star}_{MK}(\Omega) R(\Omega) \: ,
\end{equation}
$\Omega = (\alpha, \beta, \gamma)$ denoting the three Euler angles.
As the intrinsic spin of the nuclei is $0^+$ the total angular
momentum $J$ is given by the orbital angular momentum $L$ of the
relative motion of the two nuclei. As the two nuclei are identical we
will only encounter even angular momenta and positive parity states.
Because of the axial symmetry of the system only the integration over
the azimuthal angle $\beta$ has to be carried out explicitly which we
do using a Gauss-Legendre scheme with 50 points. Because of the
property
\begin{equation}
  (P^J_{MK})^\dagger P^{J'}_{M'K'} = \delta_{JJ'} \delta_{MM'}
  P^J_{KK'}
\end{equation}
we can calculate the projected GCM Hamiltonian and overlap functions as
\begin{equation}
  H^L(R,R') = \frac{4\pi}{2L+1} 
  \matrixe{\Psi(R\vec{e_z})}{(H-T_\text{cm}) P^L_{00}}{\Psi(R'\vec{e}_z)}
\end{equation}
and
\begin{equation}
  N^L(R,R') = \frac{4\pi}{2L+1} 
  \matrixe{\Psi(R\vec{e_z})}{P^L_{00}}{\Psi(R'\vec{e}_z)}
  \: .
\end{equation} 
The calculated GCM energy surfaces
\begin{equation}
  E^L(R) = \frac{H^L(R,R)}{N^L(R,R)} - (E_{b1} + E_{b2}) \: ,
\end{equation}
where $E_{b1}$ and $E_{b2}$ are the ground state energies of the fragment
nuclei, are shown for $L=0$ in Fig.~\ref{fig:gcmsurfaces}. The GCM
energy does not approach the Coulomb potential at large separations as
the kinetic energy due to localization of the nuclei is still
included. At small separations we can observe the effects of
antisymmetrization and saturation that are missing in a double-folding
potential.

We can fully describe the system of the two nuclei using the GCM basis
states Eq.~\eqref{eq:gcmbasisstate}
\begin{equation}
  \label{eq:gcmwavefunction}
  \begin{split}
    \ket{\Psi_{LM}} & = 
    \int dR \: R^2 \: P^L_{M0} \ket{\Psi(R \vec{e_z})} f(R) \\
    & \approx \sum_i P^L_{M0} \ket{\Psi(R_i \vec{e_z})} f_i \: .
  \end{split}
\end{equation}
But the GCM ``wave function'' $f(R)$ can not be interpreted as a wave
function for the relative motion of two nuclei. This is because of the
non-orthogonality of the basis states. Friedrich \cite{friedrich81}
proposed a two-step approach for dealing with the non-orthogonality.
In a first step a transformation from the GCM into the Resonating
Group Method (RGM) basis is performed. This eliminates the
non-orthogonality that is caused by the use of Slater determinants in
the GCM. In a second step the non-orthogonality due to the
antisymmetrization between the two nuclei at short distances is
removed.

\subsection{RGM wave functions}

In the RGM the wave function for the relative motion of the two nuclei
is given using basis states $\ket{\Phi(\vec{r})}$
\begin{equation}
  \braket{\vec{\rho},\xi_1,\xi_2}{\Phi(\vec{r})} = 
  \mathcal{A} \left\{ \delta(\vec{r}-\vec{\rho}) \phi(\xi_1)
    \phi(\xi_2) \right\} \: .
\end{equation}
Here $\phi(\xi)$ are the intrinsic wave functions of the nuclei. If
all the single-particle states \eqref{eq:spstate} in the GCM Slater
determinants have the same width parameter $a$ the GCM basis states
can be expressed with the RGM basis states as
\cite{horiuchi:cluster,baye77}
\begin{equation}
  \ket{\Psi(\vec{R})} = \int d^3r \: \Gamma(\vec{R}-\vec{r})
  \ket{\Phi(\vec{r})} \ket{\Phi_\text{cm}}
\end{equation}
with
\begin{equation}
  \Gamma(\vec{R} - \vec{r}) = \left( \frac{\mu}{\pi a} \right)^{3/4}
  \exp \left( -\mu \frac{(\vec{R}-\vec{r})^2}{2a} \right), 
  \quad \mu = \frac{A_1 A_2}{A_1+A_2}
\end{equation}
and the center-of-mass wave function
\begin{equation}
  \braket{\vec{X}}{\Phi_\text{cm}} = \left( \frac{A}{\pi a}
  \right)^{3/4} \exp \left( -A \frac{\vec{X}^2}{2a} \right),
\quad A = A_1+A_2 \: .
\end{equation}

We introduce the angular momentum projected basis states
\begin{equation}
  \ket{\Phi_{LM}(r)} = \sqrt{\frac{4\pi}{2L+1}} P^L_{M0}
  \ket{\Phi(r \vec{e}_z)} \: .
\end{equation}
The RGM norm kernel
\begin{equation}
  \label{eq:rgmnormkernel}
  n_L(r,r') = \braket{\Phi_{LM}(r)}{\Phi_{LM}(r')}
\end{equation}
asymptotically behaves as
\begin{equation}
  n_L(r,r') \overset{r,r' \rightarrow \infty}{=}
  \frac{1+\delta_{12}(-1)^L}{r^2} \delta(r-r') \: .
\end{equation}
In the case of identical nuclei only even $L$ are allowed but an
additional factor of two appears in the normalization (exchanging all
the nucleons between the two nuclei does not change the wave function).

The angular momentum projected GCM basis state can be expressed by the
angular momentum projected RGM basis state as
\begin{equation}
  P^L_{M0} \ket{\Psi(R \vec{e_z})} = 
  \int dr \: r^2 u_L(R;r) \ket{\Phi_{LM}(r)} \ket{\Phi_\text{cm}}
\end{equation}
with
\begin{equation}
  u_L(R;r) = \sqrt{\frac{2L+1}{4\pi}} \Gamma_L(R;r)
\end{equation}
and
\begin{equation}
  \Gamma_L(R;r) = 4\pi \left( \frac{\mu}{\pi a} \right)^{3/4} \exp \left(
    -\mu \frac{R^2+r^2}{2a} \right) i_L \left(\mu \frac{R r}{a}
  \right) \: ,
\end{equation}
where $i_L$ are the modified spherical Bessel functions
\cite{horiuchi:cluster}. 

The system of two nuclei \eqref{eq:gcmwavefunction} can now be
described using RGM basis states
\begin{equation}
  \ket{\Psi_{LM}} = \int dr \: r^2 \varphi^\mathrm{RGM}(r)
  \ket{\Phi_{LM}(r)} \ket{\Phi_\text{cm}}
\end{equation}
with the RGM wave function
\begin{equation}
  \label{eq:rgmwavefunction}
  \varphi^\mathrm{RGM}(r) = \sum u_L(R_i; r) f_i \: .
\end{equation}

\subsection{Collective wave functions}

Asymptotically the RGM wave function \eqref{eq:rgmwavefunction}
possesses the properties of a ``proper'' wave function and can for
example be used for implementing boundary conditions. At short
distances the RGM norm kernel Eq.~\eqref{eq:rgmnormkernel} deviates
from the asymptotic behavior because of the antisymmetrization between
the two nuclei. To obtain a wave function that can be identified as a
wave function for the relative motion of two point-like nuclei we have
to transform the RGM basis states $\ket{\Phi_{LM}(r)}$ into new basis
states $\ket{\tilde{\Phi}_{LM}(r)}$ that fulfill
\begin{equation}
  \braket{\tilde{\Phi}_{LM}(r)}{\tilde{\Phi}_{LM}(r')} = 
  \frac{1+\delta_{12}(-1)^L}{r^2} \delta(r-r') \: .
\end{equation}

This is achieved by
\begin{equation}
  \ket{\tilde{\Phi}_{LM}(r)} = \int dr' \: r'^2 \ket{\Phi_{LM}(r')} \:
  n_L^{-1/2}(r',r) \: .
\end{equation}
For the evaluation of the norm kernel we use its spectral
representation
\begin{equation}
  n_L(r,r') = \sum_\alpha \chi^{(\alpha)}(r) \: \mu^{(\alpha)} \:
  \chi^{(\alpha)}(r') 
\end{equation}
where $\mu^{(\alpha)}$ are the eigenvalues and $\chi^{(\alpha)}$ are
the (real) eigenvectors of the norm kernel. We perform the diagonalization
numerically using the GCM basis states. For that we expand the
eigenvectors
\begin{equation}
  \chi^{(\alpha)}(r) = \sum_i u_L(R_i;r) \: \chi^{(\alpha)}_i
\end{equation}
and transform the integral integration
\begin{equation}
  \int dr' r'^2 n_L(r,r') \chi^{(\alpha)}(r') = \mu^{(\alpha)}
  \chi^{(\alpha)}(r) 
\end{equation}
into the generalized eigenvalue problem
\begin{equation}
  \sum_j n_L^{ij} \chi^{(\alpha)}_j = \mu^{(\alpha)} \sum_j \eta_L^{ij} \chi^{(\alpha)}_j 
\end{equation}
with
\begin{equation}
  \begin{split}
    n_L^{ij} &= \int dr \: r^2 \int dr' \: r'^2 u_L(R_i;r) n_L(r,r')
    u_L(R_j;r') \\
    &= \matrixe{\Psi(R_i\vec{e_z})}{P^L_{00}}{\Psi(R_j\vec{e_z})}
  \end{split}
\end{equation}
and
\begin{equation}
  \eta_L^{ij} = \int dr \: r^2 u_L(R_i;r) u_L(R_j;r) \: .
\end{equation}
We can now express the GCM basis state as
\begin{equation}
  P^L_{M0} \ket{\Psi(R_i \vec{e_z})} = \int dr \: r^2 \tilde{u}_L(R_i; r)
  \ket{\tilde{\Phi}_{LM}(r)} \ket{\Phi_\text{cm}}
\end{equation}
with
\begin{equation}
  \begin{split}
    \tilde{u}_L(R_i; r) & = 
    \int dr' \: r'^2 n_L^{1/2}(r,r') \: u_L(R_i; r') \\
    & = \sum_{k} u_L(R_k; r) \sum_l \left(
    \sum_\alpha \chi^{(\alpha)}_k \sqrt{\mu^{(\alpha)}} \chi^{(\alpha)}_l
  \right) \eta_L^{li} \: .
  \end{split}
\end{equation}

\begin{figure}
  \includegraphics[width=\columnwidth]{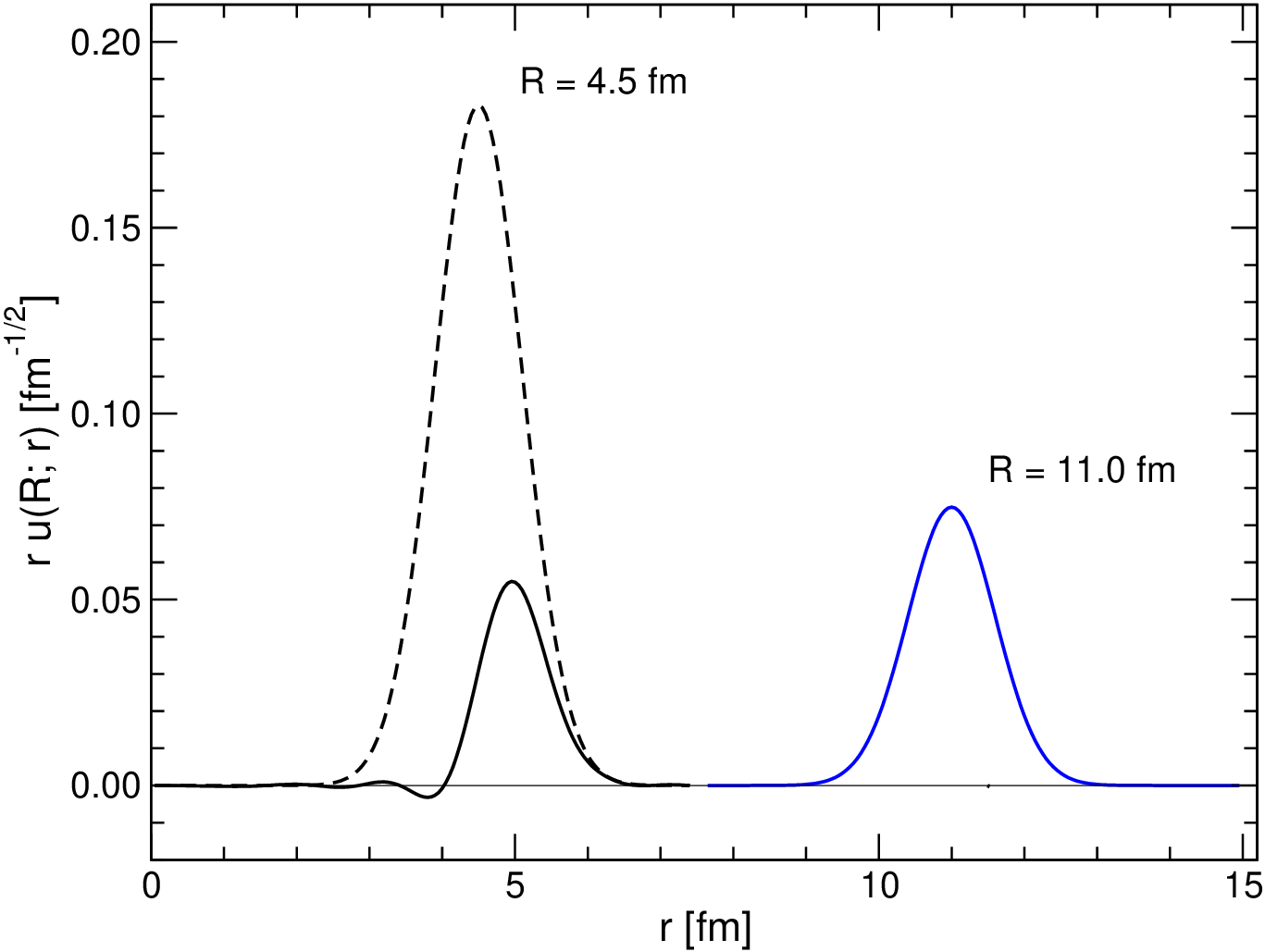}
  \caption{The wave functions $r u_L(R;r)$ (dashed lines) and $r
    \tilde{u}_L(R;r)$ (solid lines) for
    \isotope{16}{O}-\isotope{16}{O}, $L=0$ at $R=4.5\:\fm$ and
    $R=11\:\fm$. $\tilde{u}_L(R,r)$ is suppressed at small distances
    due to antisymmetrization.}
  \label{fig:basisstates}
\end{figure}

The effects of antisymmetrization are illustrated in
Fig.~\ref{fig:basisstates} where $u_L(R;r)$ and $\tilde{u}_L(R;r)$ are
shown for small and large separations. At large separations $u_L$ and
$\tilde{u}_L$ coincide while at $R=4.5\:\fm$, where there is a large
overlap between the surfaces of the nuclei, $\tilde{u}$ is
substantially reduced and shifted outwards due to antisymmetrization.
Completely Pauli forbidden states correspond to eigenvalues $\mu=0$.
For large separations the antisymmetrization between the clusters has
no effect and the eigenvalues $\mu$ approach 1.

For a state described in the GCM world as
\begin{equation}
  \ket{\Psi_{LM}} = \sum_i P^L_{M0} \ket{\Psi(R_i \vec{e_z})} f_i
\end{equation}
the corresponding proper wave function for the relative motion is
given by
\begin{equation}
  \varphi_L(r) = \sum_i \tilde{u}_L(R_i; r) f_i \: .
\end{equation}

\section{\label{sec:potentials} Nucleus-Nucleus Potentials}

In principle the transformation procedure defines an energy
independent effective interaction but it has a complicated non-local
structure. We will therefore follow Friedrich \cite{friedrich81} and
fit a local potential to the GCM matrix elements.

When comparing the GCM overlap matrix elements with the corresponding
expression in the collective world
\begin{equation}
  \matrixe{\Psi(R_i \vec{e_z})}{P^L_{00}}{\Psi(R_i \vec{e_z})} \overset{!}{=}
  \int dr \: r^2 \tilde{u}_L(R_i;r) \tilde{u}_L(R_i;r)
\end{equation}
we observe numerical deviations for $R_i \le 3.5\, \fm$ where
$\tilde{u}_L(R_i;r)$ is almost completely suppressed. As we will solve
the Schr{\"o}dinger equation with incoming wave boundary conditions,
we will need the solution of the Schr{\"o}dinger equation only from
the minimum in the potential surface outwards to calculate the barrier
penetration. We therefore fit the Hamiltonian matrix elements only for
$R_i \ge 4.0\,\fm$. Our ansatz for the effective Hamiltonian in the
collective world is given by
\begin{equation}
  H^L_\text{eff}(r) = \frac{p^2_\text{rel}}{2\mu M_N}  +
  \frac{L(L+1)}{2\mu M_N r^2} + V^L_\text{eff}(r) + V_C(r) + E_{b1} + E_{b2} 
\end{equation}
where the Coulomb interaction is that of two homogenously charged
spheres ($r_c$ is taken to be the charge radius)
\begin{equation}
  V_C(r) =
  \begin{cases}
    Z_1 Z_2 e^2 \frac{1}{r} & ; r > 2 r_c \\
    Z_1 Z_2 e^2 (3 - \frac{r^2}{r_c^2}) \frac{1}{2r_c} & ; r < 2 r_c
  \end{cases}
\end{equation}

The effective potential $V^L_\text{eff}(r)$ is parameterized as a sum
of ten Gaussians. We minimize the deviation from the GCM matrix
elements to fulfill the relation
\begin{multline}
  \matrixe{\Psi(R_i\vec{e_z})}{(H-T_\text{cm})
    P^L_{00}}{\Psi(R_i\vec{e_z})} \overset{!}{=} \\
  \int dr \: r^2 \tilde{u}_L(R_i; r) H^L_\text{eff}(r) \tilde{u}_L(R_i; r)
\end{multline}
as well as possible. We determine $V^L_\text{eff}(r)$ independently
for $L=0,2,4$ and obtain almost identical fits. Therefore a
$L$-independent $V_\text{eff}(r)$ is chosen and fitted simultaneously
to the $L=0,2,4$ matrix elements. The deviations between the collective
and GCM matrix elements are less than $50\:\mathrm{keV}$. 

\begin{figure}
  \includegraphics[width=\columnwidth]{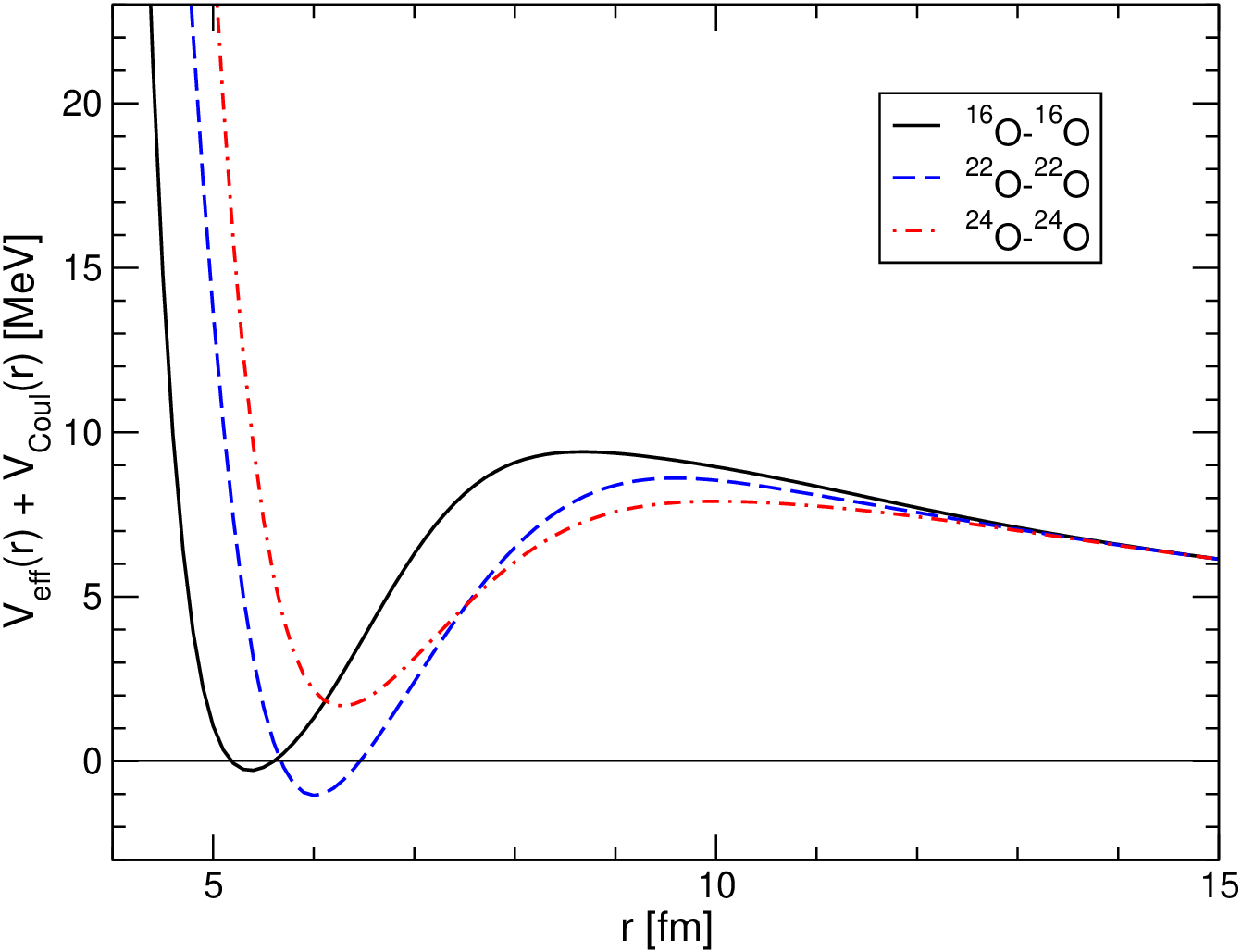}
  \caption{Effective Nucleus-Nucleus Potentials ($L=0$) derived
    from the FMD energy surfaces. The barrier gets lower and shallower
    with increased neutron number.}
  \label{fig:potentials}
\end{figure}

The fitted $L=0$ effective potentials including the Coulomb
interaction are shown in Fig.~\ref{fig:potentials}. As could already
be deduced from the GCM energy surfaces the neutron rich isotopes
feature lowered barriers at larger distances due to the attractive
nuclear interaction in the tails of the neutron distribution.

\section{\label{sec:fusion} Fusion Cross Sections}

With the effective potential derived in the last section the two-body
Schr{\"o}dinger equation 
\begin{equation}
  \left[ \frac{p^2 _\text{rel}}{2\mu M_N} + 
    \frac{L(L+1)}{2\mu M_N r^2} +
    V_\text{eff}(r) + V_C(r) \right] \varphi_L(r) = E \varphi_L(r)
\end{equation}
is solved with Incoming Wave Boundary Conditions (IWBC)
\cite{rawitscher66}. No imaginary part of the potential is needed in
this approach. We assume that the nuclei will fuse when the minimum in
the potential surface has been reached. The IWBC can be formulated as
\begin{equation}
  \frac{\varphi_L'(r_\text{min})}{\varphi_L(r_\text{min})} = 
  i k_L(r_\text{min})
\end{equation}
where $k_L(r)$ is the local wave number
\begin{equation}
  k_L(r) = \sqrt{2 \mu M_N \left(E - \frac{L(L+1)}{2\mu M_N r^2} -
    V_\text{eff}(r) - V_C(r) \right)} \; .
\end{equation}

The Schr{\"o}dinger equation is integrated numerically starting from
$r_\text{min}$ outwards and matched to the Coulomb scattering
solution
\begin{equation}
  \begin{split}
    \varphi_L^\text{Coul}(r) = \: & C_- \left(F_L(k r) - i G_L(k
      r)\right) + \\ 
    & C_+ \left( F_L(k r) + i G_L(k r) \right)
  \end{split}
\end{equation}
with
\begin{equation}
  k = \sqrt{2 \mu M_N E} \: .
\end{equation}
The matching point $r_m$ has to be outside the range of the nuclear
potential $V_\text{eff}(r)$. The penetration factor $P_L$ is given by
the ratio of the incoming flux at $r_\text{min}$ to the incoming
Coulomb flux
\begin{equation}
  P_L = \frac{k_L(r_\text{min})}{k |C_-|^2} \; .
\end{equation}

For very low energies ($E < 4\:\mathrm{MeV}$) the magnitude of the
transmission coefficient gets very small and the numerical matching is
no longer possible. For these energies we use the approximation
\cite{christensenswitkowski77}
\begin{equation}
  \label{eq:approxtransmission}
  P_L \approx \frac{k_L(r_\text{min})}{k_L(r_m)} 
  \frac{|\varphi_L(r_\text{min})|^2}{|\varphi_L(r_m)|^2}
  \frac{4 F_L^2(k r_m)}{F_L^2(k r_m) + G_L^2(k r_m)} \; .
\end{equation}

With the penetration factors we immediately obtain the fusion cross section
\begin{equation}
  \sigma(E) =  \frac{\pi}{k^2} \sum_{L=0}^{L_{\mathit{crit}}}
  (1+\delta_{12}(-1)^L) (2L+1) P_L(E) \: .
\end{equation}

As we are dealing with the fusion of identical nuclei only even
angular momenta $L$ contribute to the cross section.

\begin{figure}
  \includegraphics[width=\columnwidth]{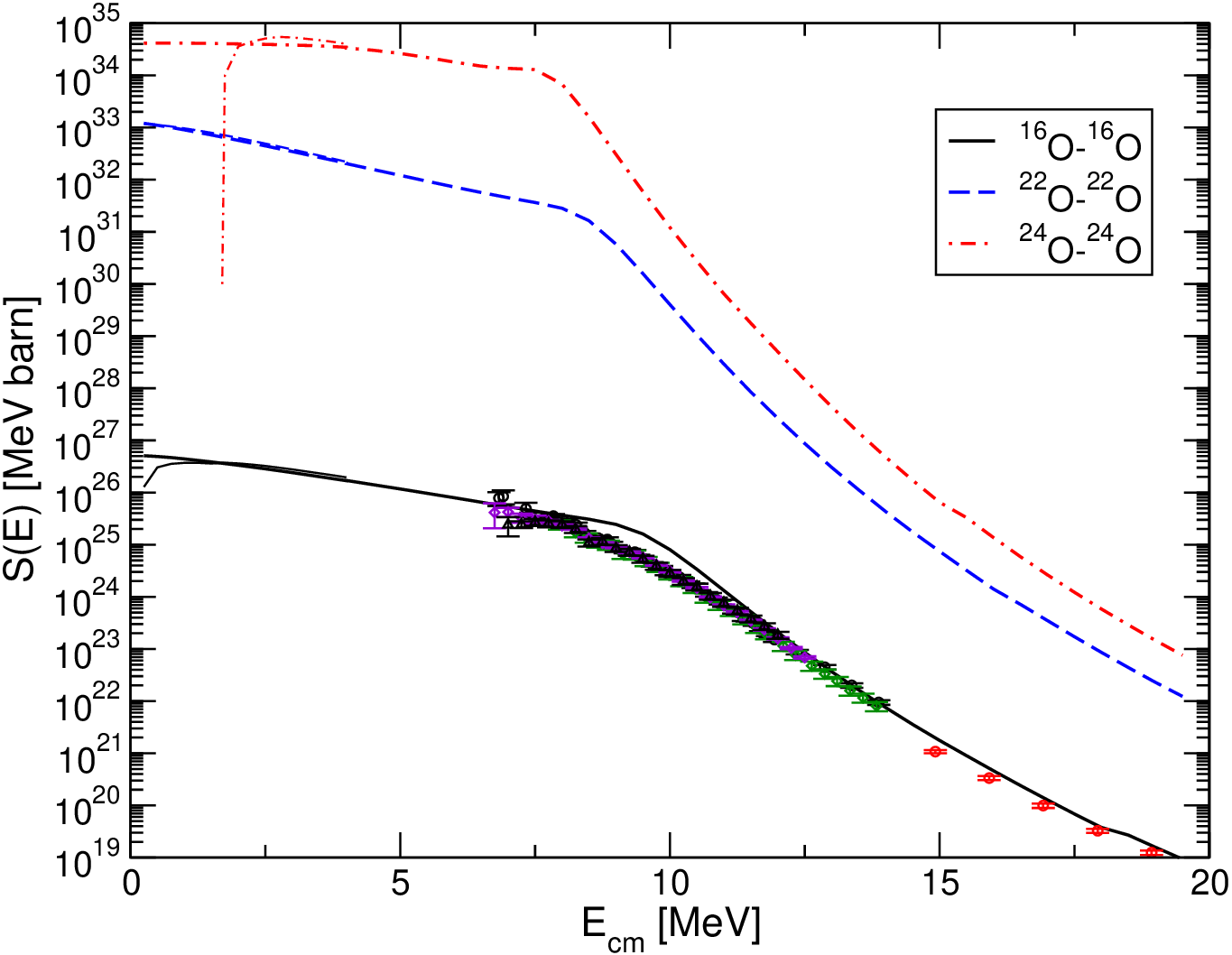}
  \caption{S-factor for the fusion of oxygen isotopes. Experimental
    data \cite{spinka72,hulke80,wu84,thomas86} for \isotope{16}{O} are
    included. The S-factor is calculated with the effective potential
    derived for the frozen configurations (thin lines) and with the
    modified potentials for the adiabatic configurations (thick
    lines).}
  \label{fig:sfactor}
\end{figure}

Instead of the cross section we plot in Fig.~\ref{fig:sfactor} the
astrophysical S-factor where the tunneling through a point-charge
$s$-wave Coulomb barrier has been taken out of the cross section
\begin{equation}
  S(E) = \sigma(E) \: E \: e^{2\pi\eta} \: .
\end{equation}

For \isotope{16}{O} the agreement with the available data is
satisfactory, taking the fact into account that no parameters have
been adjusted. The observed edge in the S-factor is related to the
barrier height (see Fig.~\ref{fig:potentials}).  In the case of
\isotope{24}{O} no fusion occurs below $E_\mathrm{cm} = 2\:\MeV$
because the potential calculated with frozen configurations does not
drop below zero inside the Coulomb barrier. In the following section
we will discuss the adiabatic energy surface for which the S-factor
will smoothly continue to $E_\mathrm{cm} = 0$.

Comparing the neutron rich oxygen isotopes with \isotope{16}{O} we
find fusion cross sections that are enhanced by 6-8 orders of
magnitude. This is caused by the reduction of the height and width of
the Coulomb barriers for the neutron rich \isotope{22}{O} and
\isotope{24}{O}.

In the last years a hindrance of the fusion process at sub-barrier
energies compared to coupled channel calculations has been found for
medium-heavy nuclei (\cite{jiang06} and references therein). In
\cite{misicu06} this effect was related to a shallower potential due
to the saturation properties of nuclear matter that become important
for large overlap of the reacting nuclei. In a recent paper
\cite{jiang07} a drop-off of the S-factor in the
\isotope{16}{O}-\isotope{16}{O} fusion reaction has been predicted for
energies below $7\:\MeV$ due to this fusion hindrance effect. Although
our wave functions are fully antisymmetrized products of frozen ground
states and therefore saturation effects are included explicitly we do
not find such a decrease in the S-factor. The nucleus-nucleus
potential while featuring only a shallow minimum still allows for an
increase in the S-factor for energies well below the barrier.

\section{\label{sec:adiabatic} Adiabatic effects}

Up to now we have discussed the fusion process in a single-channel
approximation where the nuclei stay in their ground states. To
describe effects beyond the single-channel approximation we could in
principle extend our discussion to a coupled-channel approach by
including excited states and trying to derive the coupling potentials.

\begin{figure}
  \includegraphics[width=\columnwidth]{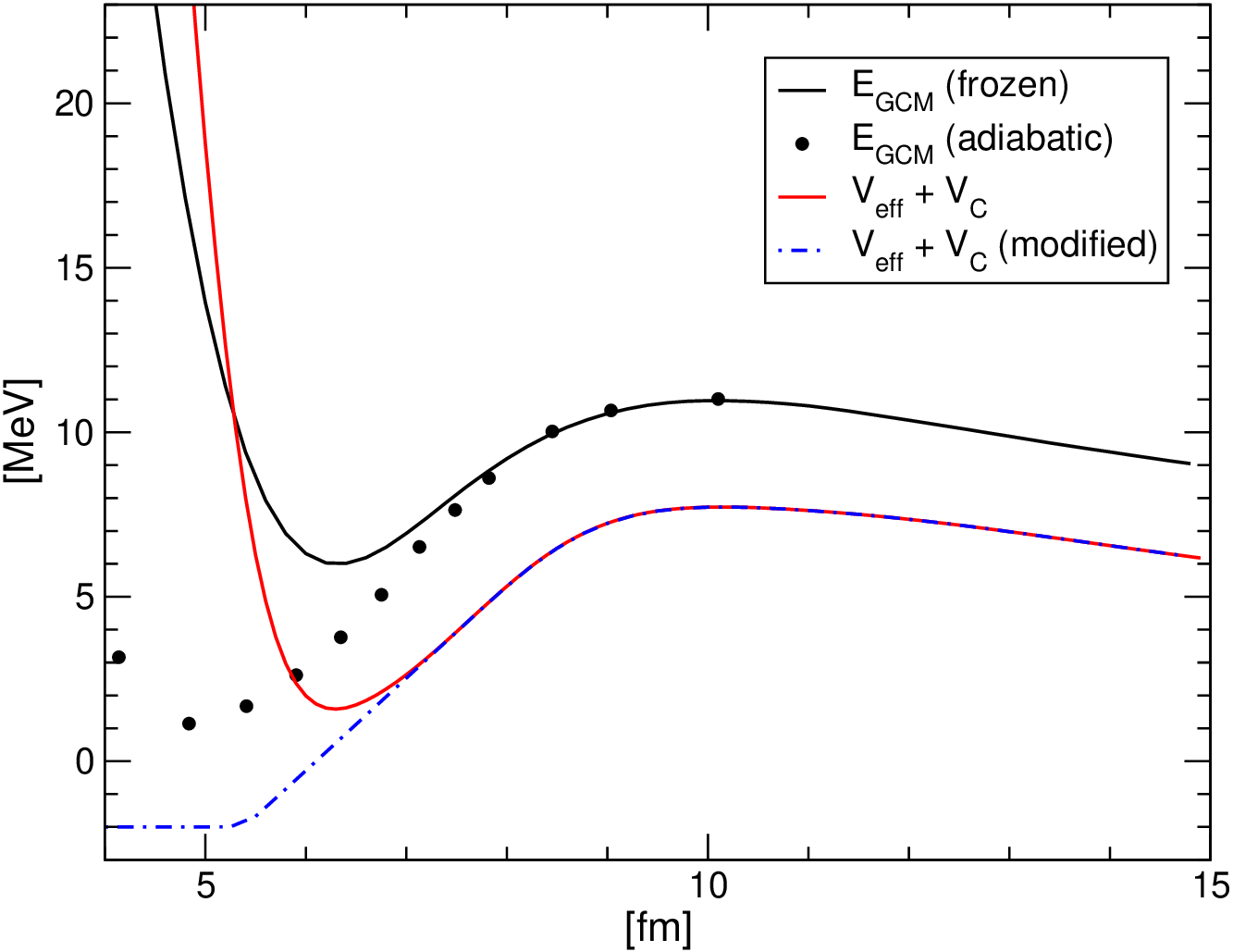}
  \caption{GCM energy surfaces for the \isotope{24}{O}-\isotope{24}{O}
    system calculated with frozen configurations (solid line) and with
    FMD configurations minimized under quadrupole constraints (dots).
    The intrinsic quadrupole deformation is mapped on the separation
    $R$ between the nuclei. The effective nucleus-nucleus potential
    derived from the frozen GCM configs is given by the dashed line.}
  \label{fig:adiabatic}
\end{figure}

Another point of view is to switch from the diabatic picture with
frozen configurations to an adiabatic approach where we allow for
distortions of the wave functions. We generate such wave functions by
minimizing the expectation value of the energy for the $2A$-nucleon
system with respect to the parameters of all the FMD single particle
states under constraints on the quadrupole deformation of the total
intrinsic wave function. For these configurations we define the
separation distance to be that of the frozen configuration with the
same quadrupole deformation. The adiabatic GCM energy surface
calculated for \isotope{24}{O} can be seen in
Fig.~\ref{fig:adiabatic}. We see almost no effect in the region of the
Coulomb barrier but observe less repulsion at smaller separations.
The corresponding adiabatic wave functions can no longer be expressed
with RGM wave functions and therefore the derivation of the effective
nucleus-nucleus potential as done for the frozen configurations is not
possible any more. To estimate the effects of the adiabatic
configurations we modify the nucleus-nucleus potential as shown in
Fig.~\ref{fig:adiabatic} to reflect the changes seen in the GCM energy
surface. These modifications change the low energy behavior especially
in the \isotope{24}{O}-\isotope{24}{O} case where the nucleus-nucleus
potential for the frozen configurations has a minimum above zero
energy. The modified potential goes below zero and the S-factor now
smoothly continues to $E_\mathrm{cm} = 0$. A similar but less
pronounced effect can be seen for \isotope{16}{O}-\isotope{16}{O},
whereas the \isotope{22}{O}-\isotope{22}{O} S-factor stays almost
unchanged. Variations of the depth of the modified potential from
$-0.5\:\MeV$ to $-10\:\MeV$ change the S-factor by less then 50\%.

\section{\label{sec:summary} Summary and Conclusion}

In this paper we studied the fusion cross sections for the oxygen
isotopes \isotope{16}{O}, \isotope{22}{O} and \isotope{24}{O} in the
Fermionic Molecular Dynamics model. The same effective nucleon-nucleon
interaction is used for the ground states and the two-nucleus system.
The total wave function is antisymmetrized and projected on angular
momentum eigenstates. In a two-step process an effective
nucleus-nucleus potential is derived from the GCM energy surface. The
Schr{\"o}dinger equation is solved with incoming wave boundary
conditions to obtain the fusion cross sections. Effects beyond the
frozen ground state approximation are studied with adiabatic
configurations obtained as FMD many-body states under constraints on
the quadrupole deformation. Experimental data on
\isotope{16}{O}-\isotope{16}{O} are reproduced; a much enhanced cross
section for the fusion of the neutron-rich oxygen isotopes is found.

\begin{acknowledgments}
  We are grateful to Michael Wiescher for suggesting to investigate
  fusion cross sections in the FMD approach and thank him and Leandro
  Gasques for fruitful discussions. T.N. acknowledges the support by
  The Joint Institute for Nuclear Astrophysics (JINA) NSF PHY 0216783.
\end{acknowledgments}


\end{document}